# The utility of wearable devices in assessing ambulatory impairments of people with multiple sclerosis in free-living conditions

Shaoxiong Sun[1], PhD; Amos A Folarin[1,2], PhD; Yuezhou Zhang[1], MSc; Nicholas Cummins[1], PhD; Shuo Liu[3], MSc; Callum Stewart[1], MSc; Yatharth Ranjan[1], MSc; Zulqarnain Rashid[1], PhD; Pauline Conde[1], BSc; Petroula Laiou[1], PhD; Heet Sankesara[1], BSc; Gloria Dalla Costa[4], MD; Letizia Leocani[4], MD, PhD; Per Soelberg Sørensen[5], MD; Melinda Magyari[5], MD, PhD; Ana Isabel Guerrero[6], MSc; Ana Zabalza[6], MD; Srinivasan Vairavan[7], PhD;  Raquel Bailon[8,9], PhD; Sara Simblett[10], PhD; Inez Myin-Germeys[11], PhD; Aki Rintala[11,12], PhD; Til Wykes[10,13], PhD;  Vaibhav A  Narayan[7], PhD; Matthew Hotopf[10,13], PhD; Giancarlo Comi[14], MD; Richard JB Dobson[1,2], PhD; RADAR-CNS consortium[15]

[1]The Department of Biostatistics and Health informatics, Institute of Psychiatry, Psychology and Neuroscience, King's College London, London, UK
[2]Institute of Health Informatics, University College London, London, UK
[3]Chair of Embedded Intelligence for Health Care & Wellbeing, University of Augsburg, Germany
[4]University Vita Salute San Raffaele, Neurorehabilitation Unit and Institute of Experimental Neurology, IRCCS Ospedale San Raffaele, Milan, Italy
[5]Danish Multiple Sclerosis Center, Department of Neurology, Copenhagen University Hospital Rigshospitalet, Copenhagen, Denmark
[6]Servei de Neurologia-Neuroimmunologia, Centre d'Esclerosi Múltiple de Catalunya (Cemcat), Vall d'Hebron Institut de Recerca, Hospital Universitari Vall d'Hebron, Universitat Autònoma de Barcelona, Barcelona, Spain
[7]Janssen Research and Development LLC, Titusville, NJ, USA
[8]Biomedical Signal Interpretation & Computational Simulation (BSICoS) Group, Aragon Institute of Engineering Research (I3A), IIS Aragon, University of Zaragoza, Zaragoza, Spain
[9]Centro de Investigacion  Biomedica en Red en Bioingeniería, Biomateriales y Nanomedicina (CIBER-BBN), Madrid, Spain
[10]Institute of Psychiatry, Psychology and Neuroscience, King's College London, London, UK
[11]Centre for Contextual Psychiatry, Department of Neurosciences, KU Leuven, Leuven, Belgium
[12]LAB University of Applied Sciences, Faculty of Social Services and Health Care, Lahti, Finland
[13]South London and Maudsley NHS Foundation Trust, London, UK
[14]Institute of Experimental Neurology, IRCCS Ospedale San Raffaele, Milan, Italy
[15]www.radar-cns.org

# Abstract

Multiple sclerosis (MS) is a progressive inflammatory and neurodegenerative disease of the central nervous system affecting over 2.5 million people globally. In-clinic six-minute walk test (6MWT) is a widely used objective measure to evaluate the progression of MS. Yet, it has limitations such as the need for a clinical visit and a proper walkway. The widespread use of wearable devices capable of depicting patients' activity profiles has the potential to assess the level of MS-induced disability in free-living conditions. In this work, we extracted 96 activity features in different temporal granularities (from minute-level to day-level) and explored their utility in estimating 6MWT scores in a European (Italy, Spain, and Denmark) MS cohort of 337 participants over an average of 10 months' duration. We combined these features with participants' demographics using three regression models including elastic net, gradient boosted trees and random forest. In addition, we quantified the individual feature's contribution using feature importance in these regression models, linear mixed-effects models, generalized estimating equations, and correlation-based feature selection (CFS). The results showed promising estimation performance with $R^2$ of 0.30, which was derived using random forest after CFS. This model was able to distinguish the participants with low disability from those with high disability. Furthermore, we observed that the minute-level ($\leq$ 8 minutes) step count, particularly those capturing the upper end of the step count distribution, had a stronger association with 6MWT. The use of a walking aid was indicative of ambulatory function measured through 6MWT. This study provides a basis for future investigation into the clinical relevance and utility of wearables in assessing MS progression in free-living conditions.

# 1. Introduction

Multiple sclerosis (MS) is a progressive inflammatory and neurodegenerative disease of the central nervous system affecting over 2.5 million people globally, and it remains a leading cause of neurological disability in young adults in developed countries [1], [2].

To evaluate the progression of MS in terms of functional, particularly ambulatory, impairments, a number of assessment criteria have been employed. Among them, the Expanded Disability Status Scale (EDSS) is the most widely used metric to quantify MS disability in neurological assessments and clinical trials [3], [4]. At the lower end of the scale (0 - 3.5), the EDSS aims to capture MS-induced impairment in eight functional systems. At the middle range (4.0 - 7.5), the EDSS focuses on impairments to walking. At the upper end of the scale (8.0 – 9.5), the EDSS is dependent upon activities of daily living. Despite its widespread applications, EDSS has been criticised for being reliant on raters' subjective examination [5]. In addition, it is unable to provide a refined granular evaluation of physical capabilities at each disability level [6].

Performance-based objective measures have emerged to alleviate the drawbacks of the EDSS [7]. The six-minute walk test (6MWT) is one of the most commonly used measures to evaluate walking speed as well as endurance and motor fatigue [8], [9]. Participants are instructed to walk back and forth in a hallway for six minutes and are allowed to rest when needed. The total distance is then measured as the 6MWT result. The 6MWT has been shown to correlate significantly with physical disability measured by EDSS [8]. Furthermore, the 6MWT has shown stronger correlations with other subjective measures of ambulation and physical fatigue than the EDSS [8]. Although the 6MWT is believed to be a reliable measure, limitations include the need for a clinical visit and a walkway with a sufficient length to allow patients to perform the test while minimizing turns [10] and patients with severe symptoms such as walking difficulty may find this test rather challenging and are unable to finish it [11]. In addition to the 6MWT, other performance-based measures have also been applied such as the 2-minute walk test (2MWT) and the timed 25 feet walk test (T25FT). The 2MWT, a shorter alternative to 6MWT, measures the distance one can walk within 2 minutes, and T25FT measures the time needed to walk 25 feet. These two tests are known to have flooring effect limitations, making them less sensitive to detect differences among patients with mild disabilities [10], [12].

The increasing availability of smartphones and wearable devices provides the opportunity to estimate the performance-based measures in free-living conditions rather than constrained clinic environments. Data from these devices could augment clinical visits, providing data with greater temporal resolution to help us to understand longitudinal disease progression, variability (particularly in relapsing-remitting MS) and execute timely interventions when needed. For instance, clinic assessments are subject to time-of-day influences such as fatigue or other activities during the day [13], [14]. Frequent evaluations of MS ambulatory impairments, which can be easily done in free-living conditions, also provide valuable information for the assessment of new treatments of MS in clinical trials [7].

Existing works have compared parameters derived from wearable devices with clinical and non-clinical measures including 2MWT [15], T25FT [16], [17], and time-up-and-go [16] as well as EDSS [18] and self-reported fatigue severity scale (FSS) [19]. Yet, very little work in the literature has compared 6MWT with wearable data [20]. In addition, the existing works collected and analysed clinical outcome measures at maximum twice at the baseline and/or at the end of the study; they did not investigate how wearable-derived parameters tracked or estimated the measures over the course of the study. In addition, they either only analysed data collected in the clinic or only extracted and compared daily step count in free-living conditions with the clinical outcome measures. As such, they did not fully explore the richness of the fine-granularity data in non-clinical settings.

In this work, we focused on and exploited the utility of wearable-derived data by extracting small epoch parameters (hour by hour or minute by minute) in free-living conditions. Furthermore, we undertook comparisons using regression models between these parameters and the 6MWT over long durations with frequently repeated measurements in a large multi-country cohort. Finally, we quantified the importance of these parameters in the regression models.

## 2. Methods and Materials

This study is part of the IMI2 RADAR-CNS major programme (radar-cns.org), which aims to evaluate remote monitoring in a range of central nervous system diseases [Major Depressive Disorder (MDD), epilepsy and Multiple Sclerosis (MS)] [21,22]. This study was co-developed with service users in our Patient Advisory Board. They were involved in the choice of measures, the timing and issues of engagement and have also been involved in developing the analysis plan and representative (s) are authors of this paper and critically reviewed it. From July 2018 to Jan 2020, 337 participants were recruited at three sites: Ospedale San Raffaele (OSR) in Milan, Italy, Centre d'Esclerosi Múltiple de Catalunya (Cemcat) at the Vall d'Hebron Institut de Recerca (VHIR) in Barcelona, Spain, and Danish Multiple Sclerosis Center (DMSC), Copenhagen University Hospital, Rigshospitalet, in Copenhagen, Denmark. These participants were all previously diagnosed with MS. Participant characteristics are described in Table 1. Out of these 337 participants, 227 had relapsing-remitting MS with subacute episodes of neurological symptoms thatr subside spontaneously to apparently normal baseline function, while the remaining 110 had secondary progressive MS which is inexorably progressive neurodegeneration typically developed after 15-25 years with the relapses [1]. Note that body mass index and MS history are missing for more than 20% and 10% of the total participants, respectively. The enrolled participants had been monitored for between 6 and 24 months. Passive data was collected using smartphones and Fitbit Charge 2/3 devices, including activity, sleep and phone usage [24]. This passive collection required no participant intervention and was implemented continuously on a 24/7 basis. In addition to the passive data, active data was collected which required clinicians and/or participants to enter data. The active data included clinician- and self-completed reports and standard walk tests, most of which were managed using Research Electronic Data Capture (REDCap) [23]. The overall open-source data collection platform (radar-base.org) has been described previously [24], and enables data to be collected, uploaded, and stored. As mentioned, we focused on how well the 6MWT reflects day-to-day activity of the study participants, as measured through wearables. A full list of the data streams collected in this study can be found in Table 2. Note that we only included data collected before Jan 22, 2020, as the pandemic induced considerable behavioural changes in the recruited participants [25].

Table 1. Participant Characteristics

|  | OSR[1] | VHIR[2] | DMSC[3] | All |
|---|---|---|---|---|
| Number of participants | 111 | 142 | 84 | 337 |
| Age when enrolled [years] | 46.2 ± 9.0[4] | 46.9 ± 9.6 | 45.2 ± 10.8 | 46.3 ± 9.7 |
| Gender (Male: Female) | 24 : 87 | 45 : 97 | 25 : 59 | 94 : 243 |
| Body mass index [kg/m²] | 23.7 ± 4.5[4] | 24.5 ± 6.0 | 25.9 ± 5.0 | 24.5 ± 5.2 |
| MS phenotype (Secondary progressive: relapsing-remitting) | 37:57 | 33:106 | 30:54 | 100:217 |
| MS history[5] [years] | 14.8 ± 8.0[4] | 14.2 ± 7.8 | 12.8 ± 7.1 | 13.9 ± 7.6 |
| EDSS when enrolled | 3.4 ± 1.3[4] | 3.4 ± 1.4 | 3.5 ± 1.3 | 3.4 ± 1.3 |
| Duration since enrolled [months] | 9.9 ± 3.5[4] | 12.7 ± 3.9 | 7.7 ± 3.8 | 10.5 ± 4.2 |

[1]Ospedale San Raffaele (OSR) in Milan, Italy
[2]Vall d'Hebron Institut de Recerca (VHIR) in Barcelona, Spain
[3]Danish Multiple Sclerosis Center (DMSC) in Copenhagen, Denmark
[4]Mean ± standard deviation
[5]MS history since first diagnosis

Table 2. Data streams used in this study

| Collection manner | Sensor modality | Data streams | Sampling duration |
|---|---|---|---|
| Passive | Fitbit | Step count | Every 1 minute |
| | Fitbit | Heart rate | 5 seconds |
| | Fitbit | Sleep stages | 30 seconds |
| Active | REDCap | 6MWT | Every three months |

In order to test for association with 6MWT, we extracted parameters from the data collected through the Fitbit devices. We first calculated intermediate parameters capturing daily activity. Then, we derived features using the statistics of these intermediate daily parameters in the 60-day time window around the clinical visit. A full list of the Fitbit-derived parameters is given in Table 3. The extraction details are given below.

The available Fitbit step count data have by default a sampling duration of 1 minute. In order to capture participants' mobility patterns at different levels of granularity, we calculated the step count sum in epochs of {1,2,3,4,5,6,7,8,9,10,11,12,30,60} minutes. The calculation was done every 1 minute, with overlapping K-1 minutes for K-minute step count sum. For example, the 10-minute step count sum was calculated with 9 minutes overlapping. Then, the maximum of each of these step count sums was determined daily starting from 6a.m. on the day until 6a.m. the next day. We also computed the daily total step count sum.

Additionally, we quantified walking intensity and endurance. For this, we computed the daily moderate walking duration where participants walked more than 82 steps in each minute [26]. We also calculated the daily maximum non-stop duration and steps where participants had consecutive minute-level non-zero step counts. Furthermore, we calculated the daily proportion of time spent in each of the four Fitbit-defined activity levels (sedentary, lightly active, fairly active, and very active) [27]. Finally, we calculated the daily mean heart rate and total sleep duration to reflect the impact of participants' activity on their physiological parameters.

When calculating these intermediate daily parameters, we only considered the data from valid days where at least 128 steps were found [16]. We studied the statistics of these daily parameters in the time window of 30 days before and 30 days after each clinical visit (excluding the visit date). The time windows were discarded for analysis if less than 6 days were valid. The statistics included the maximum, $90^{th}$ percentile, median, and interquartile range of the daily-resolution parameters. These statistics generated in the 60-day time window were used as features in the regression models and feature important quantification as discussed in the following sections. Demographic information of age, gender, need for a walking aid, and MS phenotype was also included as features in the analysis. Other demographic information was not included due to missingness.

Table 3. Fitbit-derived intermediate parameters on a daily basis. The statistics (maximum, median, $90^{th}$ percentile, and interquartile range) of these daily parameters were calculated over a 60-day period around the clinical assessment and used in the regression models and feature importance assessment.

| Feature name | Sensor modality and data stream | Elaboration |
|---|---|---|
| | | |

| Epoch (1, 2, 3, 4, 5, 6, 7, 8, 9, 10, 11, 12, 30 and 60 minutes) | Fitbit step count | The step count sum was calculated for each epoch duration (n) every 1 minute (overlapping n-1 minutes). The maximum step count sum was calculated on a daily basis. |
|---|---|---|
| Daily sum | Fitbit step count | The step count sum was calculated on a daily basis. |
| Moderate walking duration | Fitbit step count | The duration in which participants walked 82 steps per minute was calculated. |
| Maximum non-stop duration | Fitbit step count | The duration was calculated in which participants walked non-stop (non-zero step count in a single minute). |
| Maximum non-stop step count | Fitbit step count | The step count sum was calculated in which participants walked non-stop (non-zero step count in a single minute). |
| Activity level | Fitbit calories | The proportion of time spent in each of the four activity levels defined by Fitbit |
| Heart rate | Fitbit heart rate | The daily mean heart rate |
| Sleep duration | Fitbit sleep | The daily sleep duration |

We explored the utility of Fitbit-derived features in estimating 6MWT in free-living conditions. Three regression models were employed, namely random forest, gradient boosted trees, and elastic net. We chose these three models due to their robustness to multilinearity in the features, which may degrade the model performance. Random forest is a tree-based regressor, which reduces generalisation errors by adding randomisation in each split and aggregating multiple trees [28]. The gradient boosted trees produce a predictive model from an ensemble of weak predictive regression trees [29]. In each stage, a regression tree is fit on the negative gradient of the given loss function (in this work least squares). The contribution of the fitted tree to the overall regression model is shrunk with a learning rate. Elastic net is a regularized regression model striking a balance between Lasso (L1 penalty) and ridge (L2 penalty) [30]. The hyperparameters in Table 4 were tuned before using the three regression models on the test data, the split of which is given below.

Table 4. Hyperparameters to be considered in the regression models

| Hyperparameters | Considered values |
|---|---|
| **Random forest** | |
| Whether bootstrap samples are used when building trees | True, False |
| The maximum depth of the tree | 5, 10, 20, 30, 40 |
| The number of features to consider when looking for the best split | a square root, 20%, 40% of the number of features |
| The minimum number of samples required to be at a leaf node | 2, 4 |
| The minimum number of samples required to split an internal node | 5, 10 |
| The number of trees in the forest | 50, 100, 200, 400 |
| **Gradient boosted trees** | |
| The number of features to consider when looking for the best split | a square root, 20%, 40%, 50% of the number of features |
| The maximum number of terminal nodes or leaves in a tree. | 4, 6, 8 |
| The fraction of samples to be used for fitting an individual tree. | 0.5, 1.0 |
| The number of sequential trees | 100, 200, 400 |
| Learning rate | 0.005, 0.01, 0.05, 0.1 |
| **Elastic network** | |
| Alpha (Constant that multiplies the penalty terms) | 0.01,0.05,0.1,0.5,1,5,10,50,100 |
| l1_ratiofloat (The ElasticNet mixing parameter) | 0.1,0.2,0.3,0.4,0.5,0.6,0.7,0.8,0.9 |

To assess the performance of the regression models, we used 4-fold cross validation in which the data were split at the participant level. Before splitting, we shuffled participants after aggregating them. This was to ensure folds were participant independent. In each round, data from 3/4 of the participants was used for training, and 1/4 for evaluation or testing. In doing this, we ensured that the trained model saw no data from the participants held for testing. We tuned the hyperparameters on the training data using nested 4-fold cross validation, which is again split at the participant level. The cross validation was repeated 5 times (different seeds when splitting participants) to capture variance in the result. Performance in the 20 rounds (4-fold repeated 5 times) was reported using root mean square error (RMSE), median absolute error (MAE) and $R^2$. RMSE is the standard deviation of the estimation errors, penalising large errors. MAE on the other hand penalises equally the errors. $R^2$ reflects the estimation error with regard to inherent variance within the data and is used to show the reduction of variance that can be explained by the use of the regressor. When tuning the hyperparameters in the training phase, we chose MAE alone as the metric for its robustness over large errors which might arise from outliers. In evaluating the testing performance, we compared $R^2$, MAE, and RMSE with Friedman tests, respectively [31,32]. The three Friedman tests were corrected for multiple testing using the Benjamini–Hochberg procedure [33]. When a significant difference was detected, Nemenyi post hoc tests were applied for pair-wise comparisons [31]. To compare the model with full features and the model with demographic factors only, we used Wilcoxon rank-sum test. A $P<.05$ was deemed statistically significant.

In order to evaluate the relevance of the extracted features, we assessed their importance or contribution in each of the three regression models. In the random forest and gradient boosted trees, we used the built-in feature importance functionality [28,29]. In the elastic net, we first normalised the features on the training data, and applied the calculated mean and standard deviation to the testing data. The absolute coefficients associated with each independent variable (features) in the trained model were used as the feature importance. In addition, we employed linear mixed-effects models (LMEM) and generalized estimating equations (GEE) to further study feature importance. In addition to modelling cross-sectional variations, both methods are capable of handling correlations arising from repeated measurements within each participant. LMEM incorporates random components in order to adjust for the influence of a wide variety of different correlation structures existing in the repeated measures within an individual [34]. GEE allows the correlation of measures within an individual to be estimated and taken into appropriate account in the formula which generates the regression coefficients and their standard errors [34]. The relevance of features was quantified based on the test statistics (t-value) in LMEM and GEE. We reported the overall ranking of features by taking the median of the rankings derived from test statistics and feature importance. To understand the correlation structure in between features with high rankings, we calculated Pearson correlation coefficients.

We also applied a filter-based feature selection method to understand the performance of the model with a subset of features. In particular, we chose correlation-based feature selection (CFS), which maximises the correlation between features and target variables and minimises the correlation between features [35]. In this work, features were selected based on the training data in each round of cross-validation and the features that were selected over 50% of the cases were reported in the Results section. The model with CFS-selected features was compared with that with the full features using Wilcoxon rank-sum test.

To investigate the ability of the model in distinguishing high and low 6MWT scores in a cross-sectional manner, we compared the upper 25% and lower 25% of the scores (ground truth) and their corresponding estimations utilising different models. Specifically, we selected the maximum 6MWT test score (ground truth) in each participant and its corresponding estimation. The overall upper and lower 25% of the selected scores and corresponding estimation were used. The comparison was done by using Wilcoxon rank-sum test, corrected for multiple testing using Benjamini–Hochberg procedure [33]. In order to quantify the model performance in classifying upper and lower scores, we calculated area under the receiver operating characteristics curve (AUC). This work was implemented in Python 3.7.4.

## 3. Results

### 3.1 Visualisation
Figure 1 presents the distribution of 6MWT scores, number of tests and range of scores per participant in the three clinical sites. The median number of tests and median and range of 6MWT scores for each participant is 3, 400, and 40, respectively. In total, 1222 6MWT scores with valid activity (Fitbit) data were included for analysis. The completion rate for Fitbit step count data was 93.2%, which was calculated as the number of days having data over the number of days since enrolled. Figure 2 gives two examples of minute-level step count on the 7th day before the clinical visits for two participants. Compared to the participant with a 6WMT score of 135, the participant with a higher 6MWT score of 573 took more steps during the day and walked faster particularly between 9a.m. and 10a.m., and between 4p.m. and 5p.m. Figure 3 shows three scatter plots between 6WMT and an example feature (3-minute $90^{th}$ percentile) for OSR, VHIR, and DMSC, respectively. At OSR, no obvious intra-subject (longitudinal) correlation can be observed for most participants (there was little variation in scores for many participants over time), while an inter-subject (cross-sectional) correlation is visible. At VHIR, similar to OSR, the intra-participant variations in the test scores were not large and longitudinal effects were not evident. The cross-sectional relationship between 6MWT and the feature in the VHIR was weaker than OSR. At DMSC, as a result of later participant recruitment, only four participants were found to have more than four test scores. Figure 4 further shows the temporal changes in 6MWT and 3-minute $90^{th}$ percentile for three example participants with different disability levels, as seen in the different ranges of their respective 6MWT. In Figure 4 (a) - (c), we saw a general agreement in the trend seen in 6MWT and 3-minute $90^{th}$ percentile, although the timing and magnitude of changes differed.

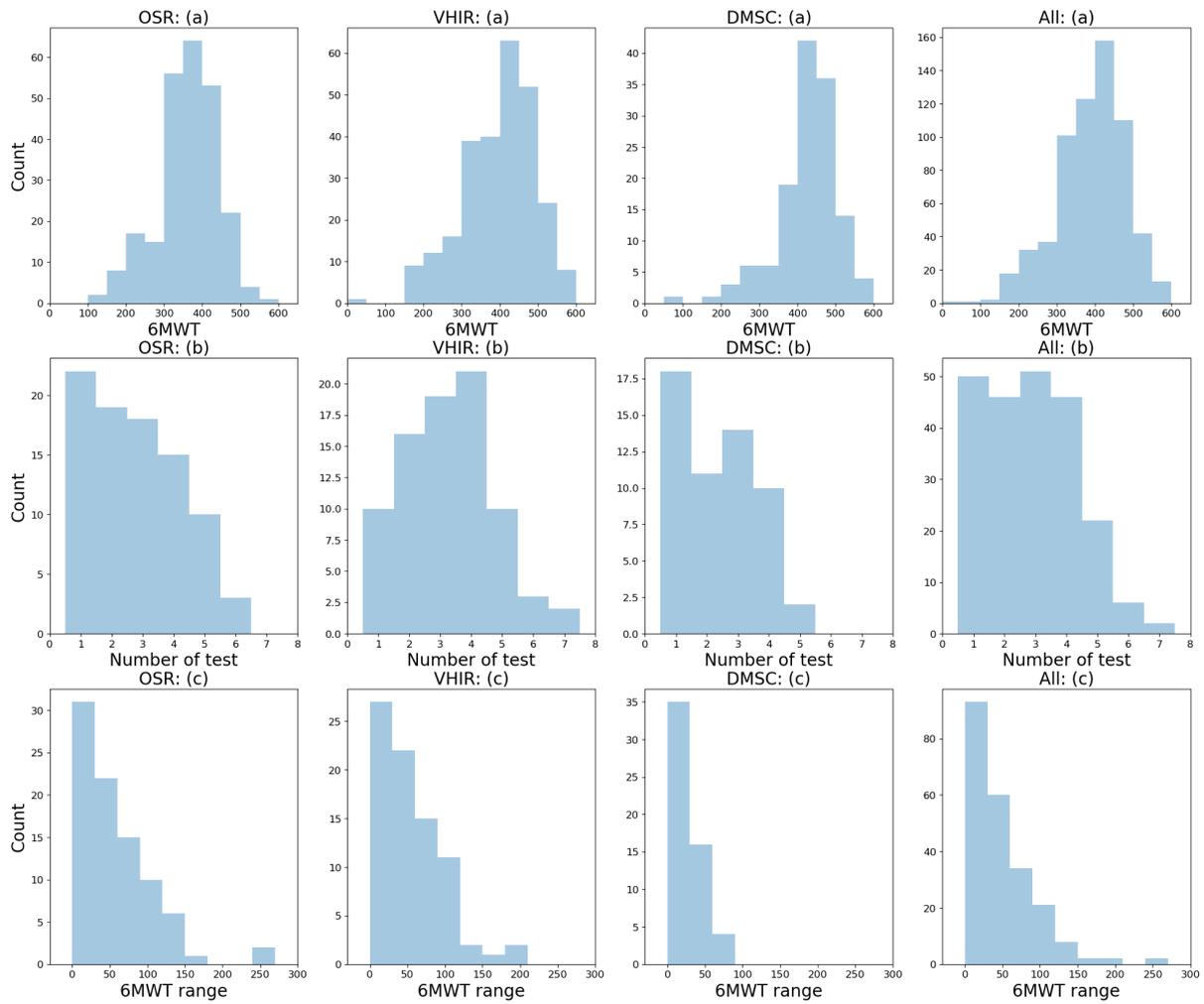

Figure 1. Histogram of 6MWT (a) scores, (b) number of tests and (c) range (maximum - minimum) by site OSR, VHIR, and DMSC.

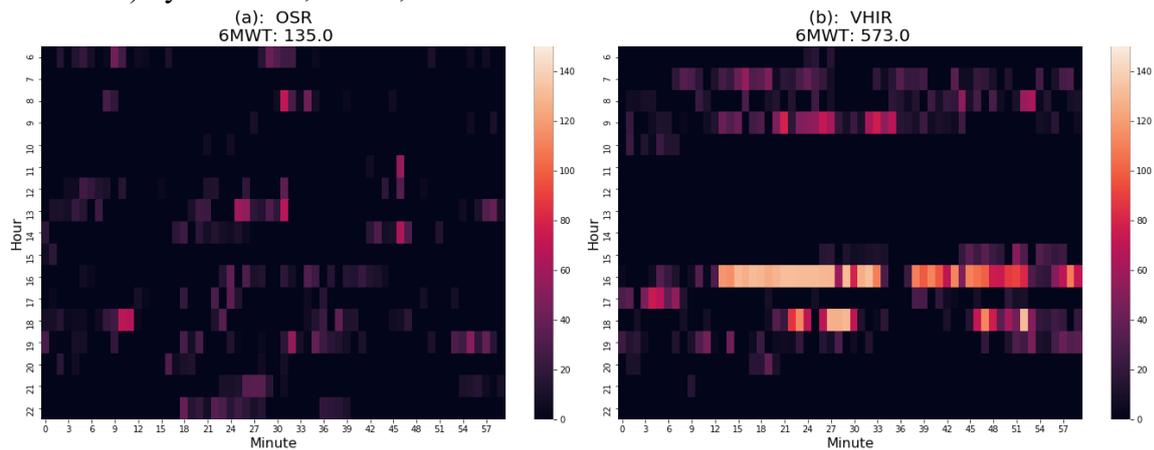

Figure 2. Minute-level step count on the 7th day before a clinical visit for 2 randomly selected participants. (a) 6MWT = 135 at the clinical site of OSR. (b) 6MWT = 573 at the clinical site of VHIR. The horizontal axis corresponds to minutes and the vertical hours. Each plot covers daytime (6a.m. to 11p.m.).

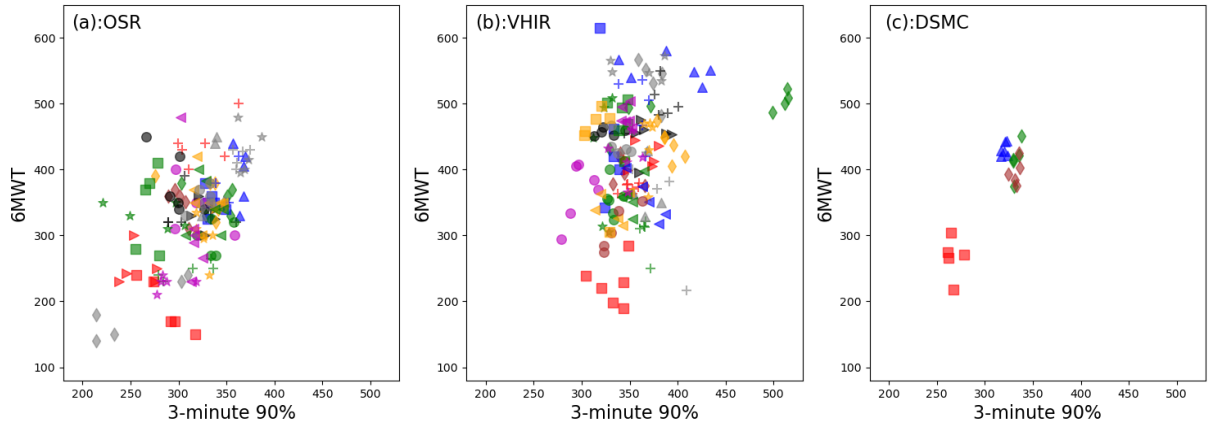

Figure 3. Scatter plots of 3-minute 90$^{th}$ percentile and 6MWT for participants with >=5 6MWT scores. (a): OSR, (b): VHIR, (c) DMSC. The markers in the same shape and color represent the same participant.

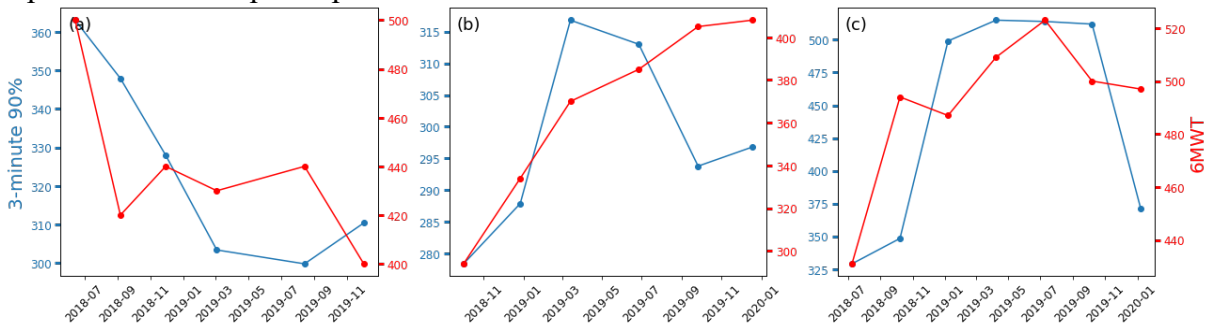

Figure 4. Temporal changes over >1 year for one Fitbit-derived feature (3-minute 90$^{th}$ percentile; blue) and 6MWT (red) in 3 (a,b,c) participants with different disability levels.

### 3.2 Regression analysis

Table 5 and Figure 5 show the estimation performance (MAE, RMSE and $R^2$) of 6MWT. These three performance indicators showed consistent results. It should be noted that the variability in the estimation performance was large across different folds in the cross-validation. We also calculated the random forest model performance with only demographic factors included (age, gender, need for a walking aid, and MS phenotype). The model with the full features had significantly lower RMSE and higher $R^2$ than that with demographic factors only.

Table 6 shows the feature coefficients, importance, and t-value in elastic net, gradient boosted trees, random forest, LMEM, and GEE for the top 20 features with the highest rankings. The rankings are further summarised and visualised in Figure 6. The rankings were generally consistent in between the three regression models and in between the two hierarchical models, while discrepancy can be observed between them as a whole. The majority of the top 20 features were the maximum and 90$^{th}$ percentile statistics of minute-level step counts. No sleep or heart rate features were seen. One activity feature (the interquartile range of the ratio of time spent in a sedentary state) and two clinical/demographic features (the use of walking aid and age) can be found in the top 20 features. Furthermore, additional contributions of the features in the presence of other features can be seen from the coefficients in the elastic net in Table 6.

While the use of walking aid had the largest absolute model coefficient, Fitbit-derived features also had large contributions. In particular, most of the high-ranking minute-level features were calculated within time windows no more than 8 minutes. Figure 7 reveals high multicollinearity in between features with high rankings, especially those with top rankings. The use of a walking aid, age, and interquartile range of sedentary duration showed moderate negative associations with the other features.

Table 5. Estimation performance (median of pooled cross-validation results)

|  | Elastic net | Gradient boosted trees | Random forest |
|---|---|---|---|
| $R^2$ | 0.27 | 0.28 | 0.25 |
| RMSE [m] | 78.7 | 77.3 | 76.7 |
| MAE [m] | 51.5 | 52.0 | 51.5 |

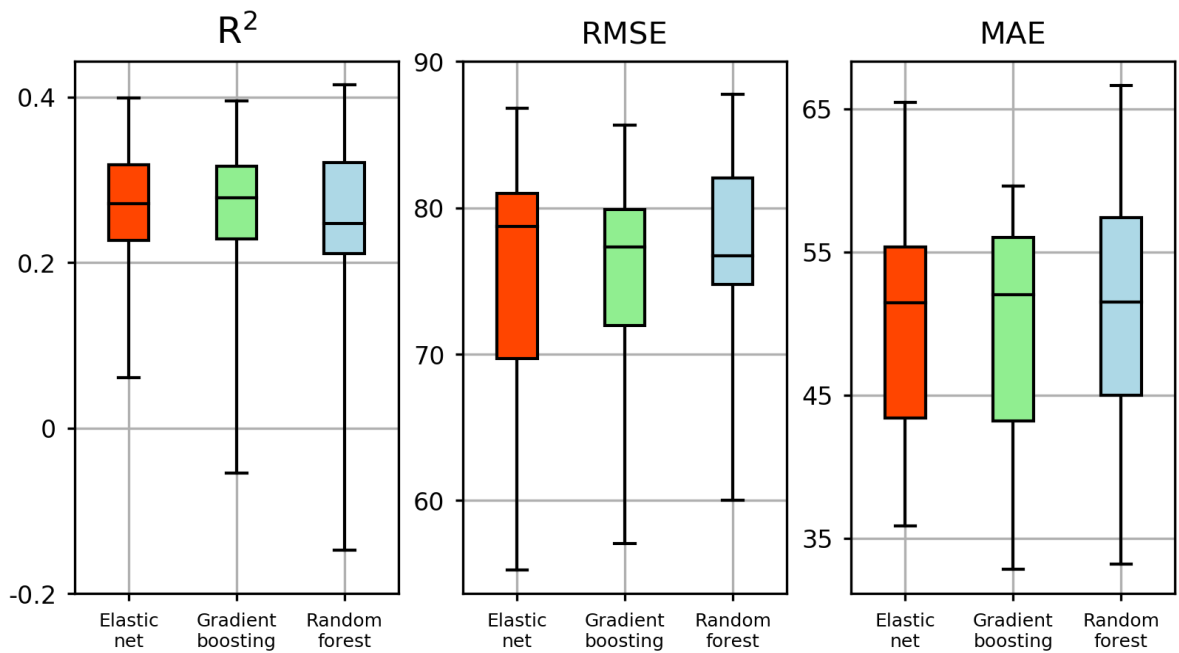

Figure 5. Estimation performance of 6-minute walk test (6MWT) scores using elastic net, gradient boosted trees, and random forest. Left: $R^2$, Centre: Root mean square errors (RMSE). Right: Median absolute error (MAE). Horizontal lines: median. Edges of boxes: 25th and 75th percentiles. Whiskers: maxima and minima.

Table 6. Top 20 Features by model contribution/importance

| Features | Median rank | EN rank | EN coef | GB rank | GB importance | RF rank | RF importance | LME rank | LME t-value | LME p-value | GEE rank | GEE t-value | GEE p-value |
|---|---|---|---|---|---|---|---|---|---|---|---|---|---|
| walking_aid | 1 | 1 | -9.86 | 1 | 0.09 | 4 | 0.04 | 1 | -7.62 | 0.00 | 27 | -5.21 | 0.00 |
| 3-minute 90th pctl | 2 | 2 | 3.52 | 2 | 0.06 | 2 | 0.04 | 7 | 6.78 | 0.00 | 16 | 6.51 | 0.00 |
| 7-minute max | 3 | 3 | 3.17 | 30 | 0.01 | 23 | 0.01 | 3 | 6.98 | 0.00 | 2 | 8.00 | 0.00 |
| 1-minute median | 5 | 5 | 1.97 | 3 | 0.06 | 3 | 0.04 | 10 | 6.67 | 0.00 | 22 | 5.49 | 0.00 |
| age | 8 | 6 | -1.90 | 5 | 0.03 | 8 | 0.02 | 61 | -2.81 | 0.00 | 67 | -2.88 | 0.00 |
| 5-minute 90th pctl | 8 | 13 | 1.40 | 8 | 0.03 | 6 | 0.03 | 8 | 6.75 | 0.00 | 11 | 6.67 | 0.00 |
| 2-minute 90th pctl | 9 | 9 | 1.72 | 4 | 0.06 | 1 | 0.05 | 23 | 5.64 | 0.00 | 24 | 5.43 | 0.00 |
| 4-minute 90th pctl | 10 | 14 | 1.32 | 11 | 0.02 | 10 | 0.02 | 5 | 6.88 | 0.00 | 8 | 6.72 | 0.00 |
| 6-minute max | 12 | 12 | 1.40 | 19 | 0.01 | 31 | 0.01 | 11 | 6.61 | 0.00 | 4 | 7.63 | 0.00 |
| 8-minute 90th pctl | 13 | 30 | 0.75 | 13 | 0.02 | 13 | 0.02 | 18 | 6.47 | 0.00 | 12 | 6.67 | 0.00 |
| 6-minute 90th pctl | 13 | 25 | 0.88 | 12 | 0.02 | 5 | 0.03 | 20 | 6.43 | 0.00 | 13 | 6.62 | 0.00 |
| 5-minute max | 14 | 15 | 1.22 | 24 | 0.01 | 11 | 0.02 | 14 | 6.51 | 0.00 | 6 | 7.22 | 0.00 |
| 7-minute 90th pctl | 16 | 18 | 1.01 | 16 | 0.01 | 9 | 0.02 | 17 | 6.48 | 0.00 | 10 | 6.67 | 0.00 |
| act-ratio-0 iqr | 16 | 11 | 1.56 | 6 | 0.03 | 16 | 0.01 | 49 | 3.71 | 0.00 | 50 | 4.10 | 0.00 |
| 8-minute max | 16 | 16 | 1.16 | 28 | 0.01 | 20 | 0.01 | 2 | 7.12 | 0.00 | 1 | 8.18 | 0.00 |
| 30-minute max | 19 | 7 | 1.83 | 9 | 0.02 | 19 | 0.01 | 26 | 5.33 | 0.00 | 25 | 5.41 | 0.00 |
| 2-minute median | 21 | 10 | 1.64 | 25 | 0.01 | 12 | 0.02 | 21 | 6.20 | 0.00 | 21 | 5.57 | 0.00 |
| 4-minute max | 22 | 27 | 0.78 | 22 | 0.01 | 18 | 0.01 | 22 | 5.97 | 0.00 | 17 | 6.47 | 0.00 |
| 3-minute max | 23 | 36 | 0.55 | 21 | 0.01 | 15 | 0.01 | 28 | 5.21 | 0.00 | 23 | 5.48 | 0.00 |
| 3-minute median | 24 | 24 | 0.91 | 14 | 0.02 | 17 | 0.01 | 24 | 5.53 | 0.00 | 33 | 4.91 | 0.00 |

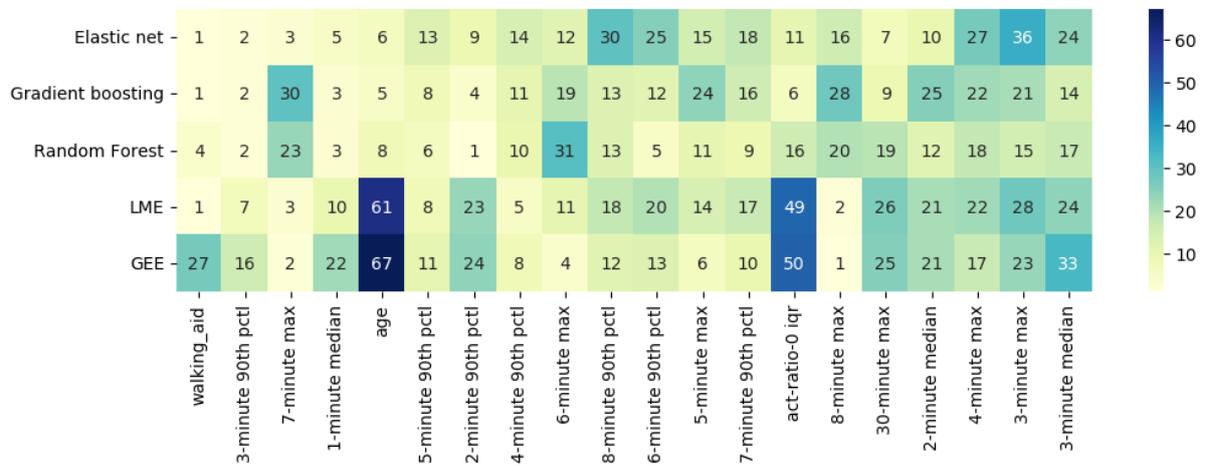

Figure 6. Feature rankings in the elastic net, gradient boosted trees, random forest, linear mixed-effects model (LMEM), and generalized estimating equations (GEE). The horizontal axis is ordered by the median rankings from each of the models. The number in each rectangle denotes the ranking of the feature in the respective model.

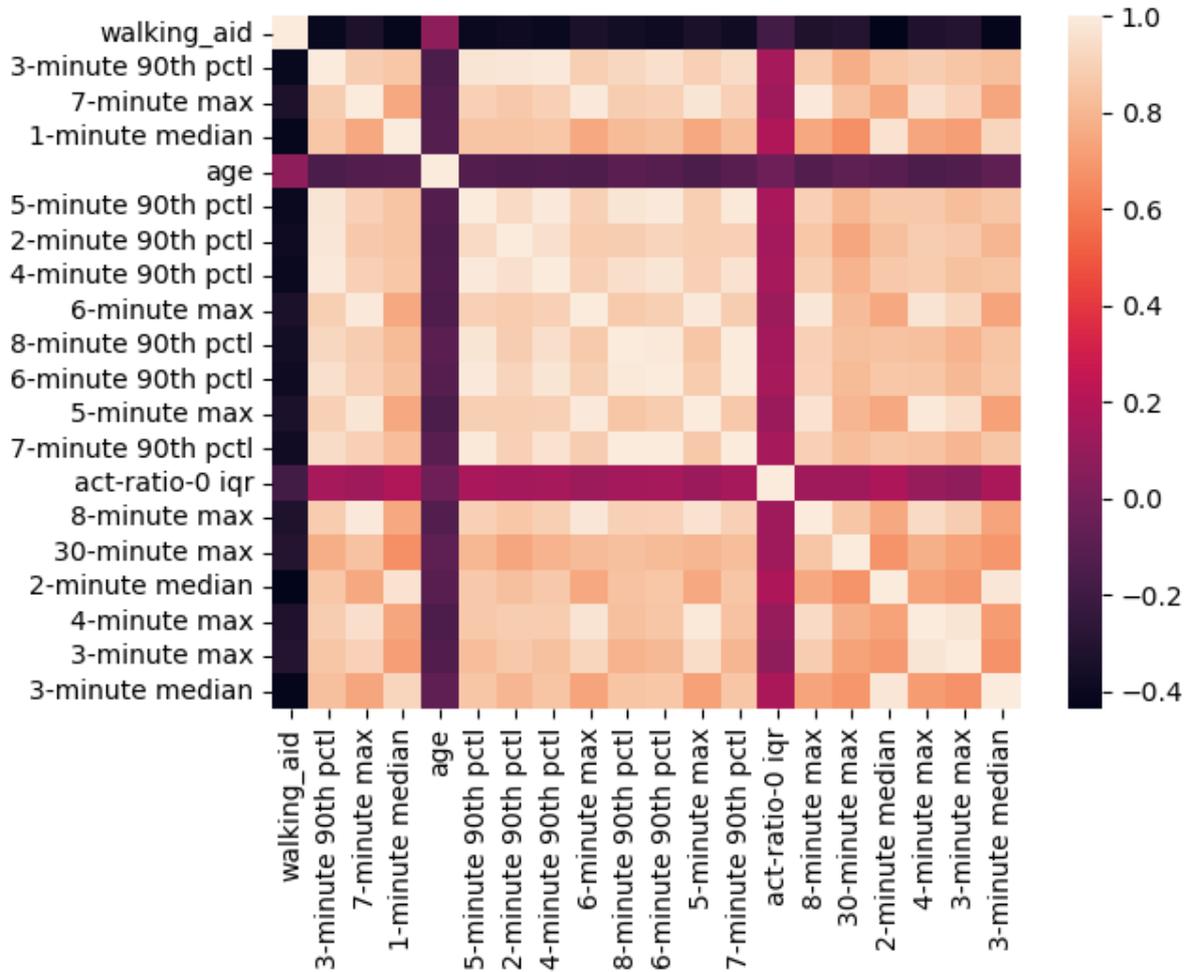

Figure 7. Pearson correlation heatmap for top 20 features (median rankings from all models)

After applying CFS, eight features were selected in over 50% of the cross-validations: 3-minute 90th percentile, the need for a walking aid, age, the proportion of time spent in the sedentary state interquartile, MS phenotype, 2-minute 90th percentile, 30-minute maximum, maximum non-stop duration interquartile. With these features, we obtained a slightly better performance, as seen in table 7 and figure 8. Yet, we did not find statistically significant difference.

Table 7. Estimation performance (median of pooled cross validation results)

|  | Elastic net | Gradient boosted trees | Random forest |
|---|---|---|---|
| $R^2$ | 0.27 | 0.29 | 0.30 |
| RMSE [m] | 76.6 | 75.5 | 76.1 |
| MAE [m] | 50.3 | 50.5 | 49.6 |

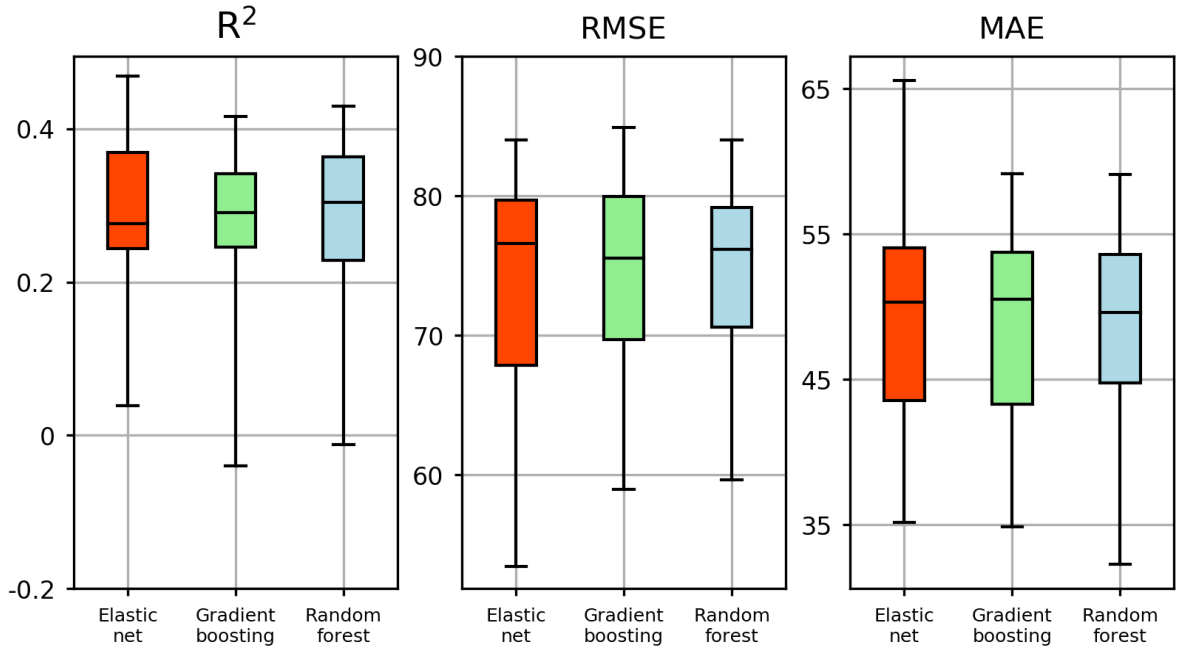

Figure 8. Estimation performance of 6-minute walk test (6MWT) scores using elastic net, gradient boosted trees, and random forest with a subset of features selected using correlation-based feature selection. Left: $R^2$, Centre: Root mean square errors (RMSE). Right: Median absolute error (MAE). Edges of boxes: 25th and 75th percentiles. Whiskers: maxima and minima.

Figure 9 shows the comparison between upper and lower 25% 6MWT scores derived from the maximum in each participant. All the models were able to show statistically significant differences between the participants with high and low 6MWT scores. The classification performance using AUC was 0.84, 0.85 and 0.87 for elastic net, gradient boosted trees and random forest, respectively.

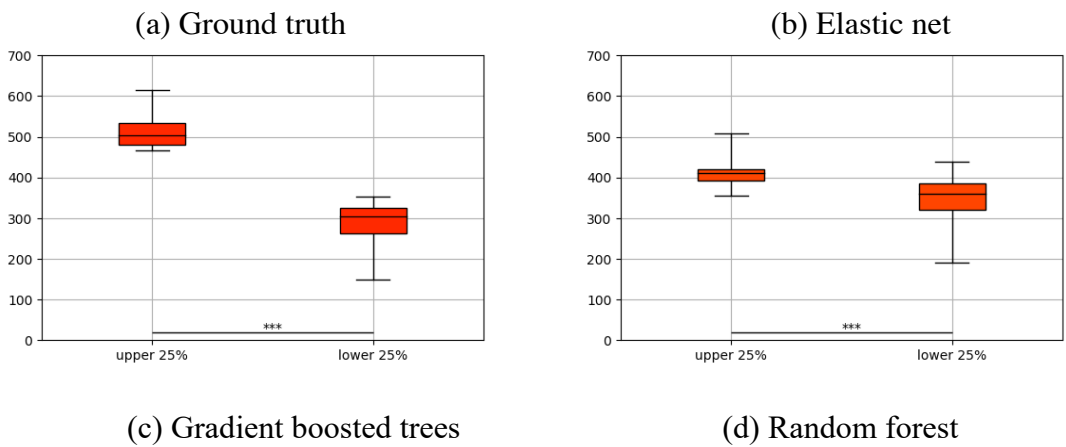

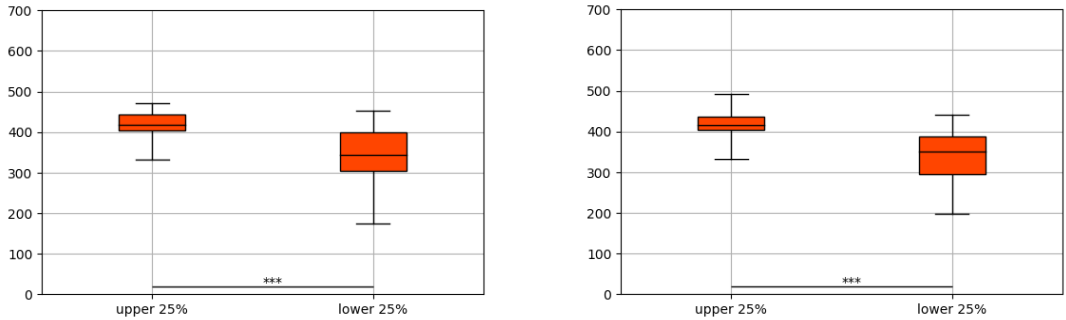

Figure 9. Comparison between upper and lower 6MWT test scores for ground truth and estimations from different models. (a) Ground truth (b) Corresponding estimation from elastic net (c) Corresponding estimation from gradient boosted trees (d) Corresponding estimation from random forest

## 4. Discussion

This study investigates the relationship between the 6MWT and parameters extracted using data collected through Fitbit wearable devices. We explored features in a wide range of temporal granularity from minute-level to daily and compared three popular regression models (elastic net, gradient boosted trees and random forest). We achieved promising estimation performance and highlighted a few features that had consistently higher contributions or more relevance in different models.

Existing works focused on the utility of daily step count when comparing the clinical test scores and Fitbit-derived passive data [16], [18]. In this study, we further examined a considerably expanding set of features in finer temporal resolution and their statistics in the time window centring the date of clinical tests. We found that the statistics of minute-level features, in particular no longer than 8 minutes, were far more predictive than those of daily features. Furthermore, among these minute-level features, it was shown that the maximum or $90^{th}$ percentile features were more strongly related to the clinical test scores than median or interquartile range features. This finding is in line with another study in which it was shown that gait speed in the standardized tests corresponds to the higher part of the distribution of the daily-life gait speed [36].

In the data visualisation, we found a stronger cross-sectional correlation between one of the most predictive features (3-minute $90^{th}$ percentile) and 6MWT across participants in comparison with the longitudinal correlation within each participant. This might be explained by observing that most participants had only been in the study less than 1.5 years, during which the disability severity was not likely to progress substantially [37]. The small variations in the measured 6MWT might be related to individual walking variability reflected in the snapshot walking test done in the clinic. Another factor affecting the relationship between 6MWT and the feature could be that step count in free-living settings may be more sensitive than 6MWT at detecting worsening ambulatory function by revealing modest early changes which are not yet captured by 6MWT [16]. Thus, the finer quantification of physical activity through wearable devices may provide a complementary or potentially a more complete view of the disease status and progression.

The 6MWT estimation performance in this study was less favourable than that in a recent study on predicting 6MWT scores on heart failure patients [38]. Among different reasons explaining the discrepancy, one plausible reason could be that people with MS often have distinctively degraded ambulation with a potential need for a walking aid; the need for a walking aid was one of the most informative features. In this study, the overall median 6MWT for participants walking freely was 410.0 metres, much higher than 264.05 metres for those in need of a walking aid. Another reason explaining the less favourable performance could be related to the longitudinal nature of this study. As discussed previously, the disability in people with MS often deteriorates gradually and remains relatively stable with natural variations over a period of one year, especially under good clinical care. This may pose challenges for estimating the repeated follow-up measures of 6MWT for some participants. Finally, digital health technologies such as Fitbit step count employed in free-living conditions might measure new constructs which the in-clinic gold standard (6MWT) does not consider, which leads to inherent discrepancy between the two [39].

In the ranking analysis, we found consistency in feature importance across different models. As discussed earlier, the maximum and $90^{th}$ percentile of minute-level step count, in particular those extracted in time windows no longer than 8 minutes, were more strongly related to 6MWT than daily features often used in the existing literature [16,18]. This finding can be explained by the observation that daily step count can be affected by other factors such as the proportion of indoor and outdoor stay and comorbidities [36]. Interestingly, the maximum step count sum in six-minute epochs in free-living conditions were also shown to differentiate between people with cardiovascular disease and controls [40]. Age had high rankings in the three regression models, possibly due to its negative correlation with 6MWT ($r = -0.13$, $p<0.05$). Interestingly, it was much less important in LMEM and GEE, which could be attributed to the use of the age at enrolment which remained the same in the subsequent repeated measurement. The reason why the extracted heart rate and sleep features were not among the top 20 may be attributed to the fact that they can be impacted by other factors such as comorbidities and the fact that we did not exploit the information contained in these two data streams. It should also be noted that the correlation between features (i.e., multicollinearity) may complicate the interpretation of feature rankings in the regression models, as shown in Figure 7. We mitigated this complication by repeating the participants' split 20 times and combining the ranking in the regression models with those in the two hierarchical models in which features were evaluated independently.

When selecting a subset of features using CFS to feed into the regression model, 6 out of the 8 most frequently selected features had high rankings. The other 2 were MS phenotype and maximum non-stop duration interquartile, which was selected later with 6 high-ranking features already in place. This suggests that these features provided complementary information to the high-ranking features. The better estimation performance using the subset of features might be explained by the use of fewer features in the model to avoid overfitting. This model with fewer features may be preferred for its computational efficiency and ease of application in a clinical setting.

To the best of the authors' knowledge, this is the first large community based longitudinal study to examine the utility of wearables in monitoring people with MS but there are some limitations of this work. First, the variation in 6MWT scores within participants was relatively small over the studied period hindering an analysis of within individual changes over time. This is most likely due to the slow developing MS-induced disability in combination with effective treatment. The ongoing RADAR-CNS study has been continuously collecting passive data, however, the outbreak of COVID-19 posed considerable challenges for the programme. As a consequence of social restrictions on mobility, participants were unable to attend review appointments to carry out 6MWT and their daily physical activity was greatly reduced [25,41].

Consequently, in this work, we did not have enough data to perform a longer analysis focusing only on the periods before the outbreak of the pandemic. Future work will attempt to find ways to incorporate the data fairly impacted by the pandemic. Second, we did not consider the possible major events of clinical relevance happening to participants during the study, which may impact the medical condition of participants. Future work will explore the complications induced by these events. Third, although the features extracted in this study cover a wide range of temporal resolutions, the description and quantification of the mobility patterns could be further extended. Future work may explore using deep learning to characterise step count profiles and find hidden patterns often unable to be captured by conventional machine learning algorithms. Fourth, we specifically excluded the data on the test date to avoid including the data during the test. If the test time slots can be recorded accurately in future works, it would be interesting to only exclude that duration and include the rest data on the day, especially before the test time slots. It could be possible that extensive travelling to clinics may also affect the test performance. In other words, we may be able to study some factors potentially causing variability in one-off measurement. Finally, while 6MWT is a widely used performance-based disability indicator, it would be also interesting in the future to see how the features and models performed on other variables such as EDSS.

## 5. Conclusion

This study demonstrated the utility of wearable Fitbit data in estimating 6MWT for people with MS in multi-country cohorts in both cross-sectional and longitudinal manners. Using Fitbit-derived features extracted in different temporal granularity, we achieved comparably promising performance with elastic net, gradient boosted trees and random forest. We also found consistency in feature importance in the three regression models and hierarchical models (LMEM and GEE). The minute-level step count, particularly those capturing the maximum or 90th percentiles of the distribution, were found to have a stronger association with 6MWT. The favourable length of the time window for calculating the step count features is generally less than or equal to 8 minutes. The use of walking aid is indicative of ambulatory function measured through 6MWT. An automatically selected subset of features may further improve the model performance. This model was able to distinguish the participants with low performances from those with high performances. This study provides a basis for future investigation into the clinical relevance and utility of Fitbit-derived parameters derived in free-living conditions.

## Acknowledgements

The RADAR-CNS project has received funding from the Innovative Medicines Initiative 2 Joint Undertaking under grant agreement No 115902. This Joint Undertaking receives support from the European Union's Horizon 2020 research and innovation programme and EFPIA, www.imi.europa.eu. This paper reflects the views of the RADAR-CNS consortium and neither IMI nor the European Union and EFPIA are liable for any use that may be made of the information contained herein. The funding body have not been involved in the design of the study, the collection or analysis of data, or the interpretation of data.

This paper represents independent research part-funded by the National Institute for Health Research (NIHR) Biomedical Research Centre at South London and Maudsley NHS Foundation Trust and King's College London. The views expressed are those of the authors and not necessarily those of the NHS, the NIHR or the Department of Health and Social Care. We thank all the members of the RADAR-CNS patient advisory board for their contribution to the device selection procedures, and their invaluable advice throughout the study protocol design.

This research was reviewed by a team with experience of mental health problems and their careers who have been specially trained to advise on research proposals and documentation through the Feasibility and Acceptability Support Team for Researchers (FAST-R): a free, confidential service in England provided by the National Institute for Health Research Maudsley Biomedical Research Centre via King's College London and South London and Maudsley NHS Foundation Trust.

RJBD is supported by the following: (1) NIHR Biomedical Research Centre at South London and Maudsley NHS Foundation Trust and King's College London, London, UK; (2) Health Data Research UK, which is funded by the UK Medical Research Council, Engineering and Physical Sciences Research Council, Economic and Social Research Council, Department of Health and Social Care (England), Chief Scientist Office of the Scottish Government Health and Social Care Directorates, Health and Social Care Research and Development Division (Welsh Government), Public Health Agency (Northern Ireland), British Heart Foundation and Wellcome Trust; (3) The BigData@Heart Consortium, funded by the Innovative Medicines Initiative-2 Joint Undertaking under grant agreement No. 116074. This Joint Undertaking receives support from the European Union's Horizon 2020 research and innovation programme and EFPIA; it is chaired by DE Grobbee and SD Anker, partnering with 20 academic and industry partners and ESC; (4) the National Institute for Health Research University College London Hospitals Biomedical Research Centre; (5) the National Institute for Health Research (NIHR) Biomedical Research Centre at South London and Maudsley NHS Foundation Trust and King's College London; (6) the UK Research and Innovation London Medical Imaging & Artificial Intelligence Centre for Value Based Healthcare; (7) the National Institute for Health Research (NIHR) Applied Research Collaboration South London (NIHR ARC South London) at King's College Hospital NHS Foundation Trust.

## Authors' Contributions

SS (Shaoxiong Sun), AAF, and RJBD contributed to the study design. SS (Shaoxiong Sun) contributed to the data analysis, figures drawing, and manuscript writing. AAF, YZ, NC, SL, PL, SV, RB, VAN, GC, MH, and RJBD contributed to the critical revision of the manuscript. AAF, YR, ZR, PC, CS, HS, and RJBD contributed to the platform design and implementation. AAF, SS (Sara Simblett), IMG, AR, TW, VAN, MH, GC, and RJBD contributed to the administrative, technical, and clinical support of the study. GDC, SS, LL, PSS, MM, AIG, and AZ contributed to data collection.

## Conflicts of Interest

# References

1. Friese, M. A., Schattling, B. & Fugger, L. Mechanisms of neurodegeneration and axonal dysfunction in multiple sclerosis. *Nat. Rev. Neurol.* **10**, 225–38 (2014).
2. Rolak, L. A. Multiple sclerosis: it's not the disease you thought it was. *Clin. Med. Res.* **1**, 57–60 (2003).
3. Kurtzke, J. F. Rating neurologic impairment in multiple sclerosis: An expanded disability status scale (EDSS). *Neurology* **33**, 1444–52 (1983).
4. Polman, C. H. *et al*. A randomized, placebo-controlled trial of natalizumab for relapsing multiple sclerosis. *N. Engl. J. Med.* **354**, 899–910 (2006).
5. Verdier-Taillefer MH, Zuber M, Lyon-Caen O, Clanet M, Gout O, Louis C, A. A. Observer disagreement in rating neurologic impairment in multiple sclerosis: facts and consequences. *Eur Neurol.* **31**, 117–9 (1991).
6. Hobart, J. Kurtzke scales revisited: the application of psychometric methods to clinical intuition. *Brain* **123**, 1027–40 (2000).
7. Goldman, M. D., Motl, R. W. & Rudick, R. A. Possible clinical outcome measures for clinical trials in patients with multiple sclerosis. *Ther. Adv. Neurol. Disord.* **3**, 229–39 (2010).
8. Goldman, M. D., Marrie, R. A. & Cohen, J. A. Evaluation of the six-minute walk in multiple sclerosis subjects and healthy controls. *Mult Scler.* **14**, 383–90 (2008).
9. Enright, P. L. *et al*. The 6-min walk test: A quick measure of functional status in elderly adults. *Chest* **123**, 387–98 (2003).
10. Bethoux, F. & Bennett, S. Evaluating Walking in Patients with Multiple Sclerosis. *Int. J. MS Care* **13**, 4–14 (2011).
11. Gijbels, D. *et al*. Predicting habitual walking performance in multiple sclerosis: Relevance of capacity and self-report measures. *Mult. Scler.* **16**, 618–26 (2010).
12. Scalzitti, D. A. *et al*. Validation of the 2-Minute Walk Test with the 6-Minute Walk Test and Other Functional Measures in Persons with Multiple Sclerosis David. *Int J MS Care.* **20**, 158–63 (2018).
13. Albrecht, H. *et al*. Day-to-day variability of maximum walking distance in MS patients can mislead to relevant changes in the expanded disability status scale (EDSS): Average walking speed is a more constant parameter. *Mult Scler.* **7**, 105–9 (2001).
14. Motl, R. W. *et al*. Validity of the timed 25-foot walk as an ambulatory performance outcome measure for multiple sclerosis. *Mult. Scler.* **23**, 704–10 (2017).
15. Paul, S. S. *et al*. Validity of the Fitbit activity tracker for measuring steps in community-dwelling older adults. *BMJ Open Sport Exerc. Med.* **1**, 1–5 (2015).
16. Block, V. J. *et al*. Association of Continuous Assessment of Step Count by Remote Monitoring With Disability Progression Among Adults With Multiple Sclerosis. *JAMA Netw. Open* **2**, 1–15 (2019).
17. Supratak, A., Datta, G., Gafson, A. R., Nicholas, R. & Guo, Y. Remote Monitoring in the Home Validates Clinical Gait Measures for Multiple Sclerosis. *Front Neurol.* **9**, 1–9 (2018).

18. Block, V. J. *et al*. Continuous daily assessment of multiple sclerosis disability using remote step count monitoring. *J Neurol* **264**, 316–26 (2017).
19. Tong, C., Craner, M., Vegreville, M. & Lane, N. D. Tracking Fatigue and Health State in Multiple Sclerosis Patients Using Connnected Wellness Devices. in *Proceedings of the ACM on Interactive, Mobile, Wearable and Ubiquitous Technologies* **3**, (2019).
20. Alexander, S., Peryer, G., Gray, E., Barkhof, F. & Chataway, J. Wearable technologies to measure clinical outcomes in multiple sclerosis: A scoping review. *Mult. Scler. J.* **27**, 1643–1656 (2021).
21. Matcham, F. *et al*. Remote assessment of disease and relapse in major depressive disorder (RADAR-MDD): a multi-centre prospective cohort study protocol. *BMC Psychiatry* **19**, 72 (2019).
22. Zhang, Y. *et al*. Relationship between major depression symptom severity and sleep collected using a wristband wearable device: Multicenter longitudinal observational study. *JMIR mHealth uHealth* **9**, 1–15 (2021).
23. https://www.project-redcap.org/
24. Ranjan, Y. *et al*. RADAR-base: Open source mobile health platform for collecting, monitoring, and analyzing data using sensors, wearables, and mobile devices. *JMIR Mhealth Uhealth*. **7**, e11734 (2019).
25. Sun, S. *et al*. Using smartphones and wearable devices to monitor behavioral changes during COVID-19. *J. Med. Internet Res.* **22**, 1–19 (2020).
26. Neven, A., Vanderstraeten, A., Janssens, D., Wets, G. & Feys, P. Understanding walking activity in multiple sclerosis: step count, walking intensity and uninterrupted walking activity duration related to degree of disability. *Neurol. Sci.* **37**, 1483–1490 (2016).
27. https://github.com/RADAR-base/RADAR-Schemas/blob/master/commons/connector/fitbit/fitbit_intraday_calories.avsc
28. Liaw, A. & Wiener, M. Classification and Regression by randomForest. *R News* (2002).
29. Friedman, J. H. Greedy function approximation: A gradient boosting machine. *Ann. Stat.* **29**, 1189–1232 (2001).
30. Zou, H. & Hastie, T. Regularization and variable selection via the elastic net. *J. R. Stat. Soc. Ser. B Stat. Methodol.* **67**, 301–20 (2005).
31. Pohlert, T. The Pairwise Multiple Comparison of Mean Ranks Package (PMCMR). *R Packag*. 27 (2014).
32. Sun S, Peeters WH, Bezemer R, Long X, Paulussen I, Aarts RM, N. G. On algorithms for calculating arterial pulse pressure variation during major surgery. *Physiol Meas* **38**, 2101–21 (2017).
33. Ferreira, J. A. & Zwinderman, A. H. On the Benjamini-Hochberg method. *Ann. Stat.* **34**, 1827–49 (2006).
34. Burton, P., Gurrin, L. & Sly, P. Extending the Simple Linear Regression Model to Account for Correlated Responses: An Introduction to Generalized Estimating Equations and Multi-Level Mixed Modelling. *Stat Med* **17**, 1261–91 (1998).
35. Hall, M. Correlation-based Feature Selection for Machine Learning. (The University of Waikato, 1999). doi:10.1.1.149.3848
36. Van Ancum, J. M. *et al*. Gait speed assessed by a 4-m walk test is not representative of daily-life gait speed in community-dwelling adults. *Maturitas* **121**, 28–34 (2019).
37. Tremlett, H., Paty, D. & Devonshire, V. Disability progression in multiple sclerosis is slower than previously reported. *Neurology* **66**, 172–7 (2006).
38. Schubert, C. *et al*. Wearable devices can predict the outcome of standardized 6-minute walk tests in heart disease. *npj Digit. Med.* **3**, (2020).

39. Mantua, V., Arango, C., Balabanov, P. & Butlen-Ducuing, F. Digital health technologies in clinical trials for central nervous system drugs: an EU regulatory perspective. *Nat. Rev. Drug Discov.* (2021). doi:10.1038/d41573-020-00168-z
40. Sokas, D. *et al.* Detection of Walk Tests in Free-Living Activities Using a Wrist-Worn Device. *Front. Physiol.* **12**, 1–13 (2021).
41. Snoeijer, B. T., Burger, M., Sun, S., Dobson, R. J. B. & Folarin, A. A. Measuring the effect of Non-Pharmaceutical Interventions (NPIs) on mobility during the COVID-19 pandemic using global mobility data. *npj Digit. Med.* **4**, 1–12 (2021).